\documentclass[fleqn,twoside]{article}
\usepackage{espcrc2}

\usepackage{graphicx}
\usepackage[figuresright]{rotating}

\newcommand{\AmS}{{\protect\the\textfont2
  A\kern-.1667em\lower.5ex\hbox{M}\kern-.125emS}}

\title{Improvement and Taste Symmetry Breaking for Staggered Quarks
\thanks{Talk presented by E. Follana. Work supported by PPARC.}  
\thanks{The simulations for
this work were carried out on a computer network running Condor at
Dallas Southern Methodist University.  The Condor Software Program
(Condor) was developed by the Condor Team at the Computer Sciences
Department of the University of Wisconsin-Madison. All rights, title,
and interest in Condor are owned by the Condor Team. \newline  
We thank K. Hornbostel for his useful help in connection with this work. \newline
We thank Anna Hasenfratz for her assistance in implementing the HYP action.}
}

\author{E. Follana \address{Department of Physics and Astronomy, 
        University of Glasgow, Glasgow G12 8QQ, UK.},
        C. Davies \addressmark,
	A. Hart \address{School of Physics, 
	  University of Edinburgh, Edinburgh EH9 3JZ, UK.},
	P. Lepage \address{Newman Laboratory for Elementary-Particle Physics,
	Cornell University, Ithaca NY 14853, USA.},
	Q. Mason \addressmark,
	H. Trottier \address{Department of Physics, Simon Fraser University.	
	Burnaby, British Columbia V5A 1S6, CANADA.}, 
	HPQCD and UKQCD collaborations.
}
       
\begin{document}

\begin{abstract}

We compare several improved actions for staggered quarks. We study the
effect of improvement on the taste changing interactions by
calculating the splitting in the pion spectrum. We investigate the
effect of the improvement on some topological properties.

\end{abstract}

\maketitle

\section{Introduction}

It is well known that the naive discretization of quark fields in
lattice QCD results in the generation, in dimension $d$, of $2^d$
species of fermions, which we will call `tastes'. The theory also
contains taste-changing interactions. This implies, for example, that
the spectrum of the theory will have several non equivalent versions
of the usual hadrons.

The staggered fermion formulation, in 4 dimensions, reduces the number
of doublers to 4. The taste-changing interactions are perturbative,
and can be reduced by improving the action
\cite{Lepage,MILC1,MILC2,Quentin}. The breaking of taste symmetry is
most clearly seen in the splitting of the pion spectrum. In the
staggered formulation there are 16 different pions, which are grouped
in 5 (approximate) multiplets. Calculating the splitting between the
pion masses in the different multiplets gives a measure of the
breaking of taste symmetry.

\section{Improved Staggered Actions and Taste Symmetry Breaking}

We have investigated the effect of taste symmetry breaking
interactions for several improved staggered action, some of them
inspired by perturbative calculations, others by perfect action ideas.

First we have the asqtad action, which removes taste-changing
interactions at tree level \cite{Lepage,MILC1,MILC2} and is fully
$a^2$ improved. If we don't include the Lepage and Naik terms, we
obtain the so-called fat7 action, which can be viewed as the
unimproved staggered action calculated on fattened (and tadpole
improved) links. We can iterate this procedure, using the fattened
links obtained in a step of fat7 to build asqtad or fat7 again (and
for these we use the obvious notation fat7xfat7, etc.) We also have
the choice of projecting the fattened links back to $SU(3)$ or not.

Another action which we study is the so called hypercubic blocked
(HYP) action, which can be viewed as the Kogut-Susskind action built
on a different kind of fattened links \cite{Anna}.

We have used a set of $\approx 200$ quenched Wilson pure gauge UKQCD
configurations, with a lattice size of $16^3 \times 32$, at $\beta =
5.9$ and $a \approx 0.1$ fm. With these configurations we do a
quenched calculation to obtain the complete set of pions for each of
the actions previously described. The fitting of pion correlators is
done using a Bayesian fitting method.

In figure 1 we plot the quadratic mass deviation from the Goldstone
mass in lattice units, $\Delta m_\pi^2a^2$, for the three central
multiplets. We see that both the HYP action and the iterated fattened
actions result in a considerable reduction of the splitting with
respect to the asqtad or fat7 action. Furthermore, it has been noticed
before \cite{MILC2} that the reunitarization of the fattened links has
a considerable effect on the splitting. Our results agree with this
\cite{Our paper}.

\begin{figure}
\includegraphics[scale=0.4, angle=-90]{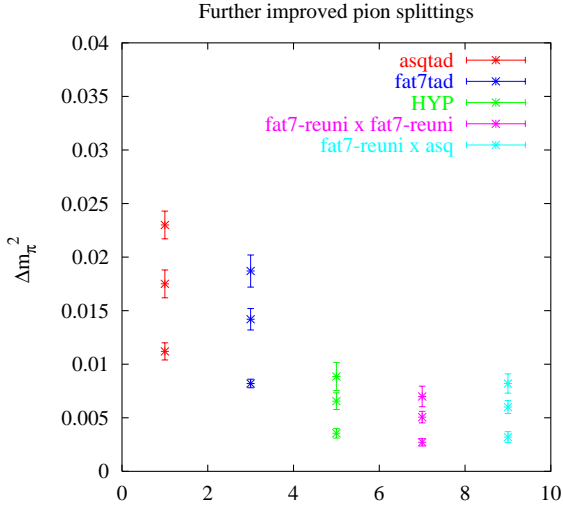}
\label{fig1}
\caption{Pion splitting: mass squared difference between central
multiplets and Goldstone pion, for several improved staggered actions}
\label{pion1}
\end{figure}

\section{Improved Staggered Actions and Topology}

The usual gluonic definition of topological charge is given by

\begin{equation}
\nonumber 
Q = {1 \over 32 \pi^2} \sum_x {1\over 32} \epsilon_{\mu \nu \sigma\tau}^{\pm} 
Tr U_{\mu \nu} (x) U_{\sigma \tau} (x)
\end{equation}

whereas the fermionic definition is 

\begin{equation}
\nonumber
Q  =  {m \over n_f} \, \, tr(\gamma_5 S_F) = 
{m^2 \over n_f} \sum_n \frac{<n|\gamma_5|n>}{\lambda^2 + m^2}
\end{equation}

where $|n>$ are the eigenfunctions of the Dirac operator in the given
gauge field background \cite{Vink}.

In the continuum the index theorem holds, and we have $Q = n^+ - n^-$,
where $n^+$ is the number of positive chirality zero modes, and $n^-$
the number of negative chirality ones. In the lattice, for a
non-chiral discretization of the Dirac operator, we lose the index
theorem. However, we can still hope to find a correlation between
``approximate zero modes'', which have a high value of the chirality
$<n|\gamma_5|n>$, and gluonic topological charge.

In our set of quenched configurations, we first calculate the
topological charge using the gluonic definition after a cooling
procedure. We also calculate the 20 lowest lying eigenvectors for both
the unimproved staggered and for several of the improved actions. In
figure 2 we show the chirality for the first 20 eigenvalues of the
Dirac operators in a configuration where, should the index theorem
hold, the first 12 eigenvectors would have been chiral. We can see a
clear qualitative difference as we improve the action, with the
appearance of approximate zero modes in the required number needed to
produce the given topological charge.


To get a more quantitative measure, we calculate $\sum_n
<n|\gamma_5|n>$ over the 20 eigenvectors. If the lowest eigenmodes are
approximately chiral (and small), and the rest are not, this quantity
should be approximately equal to the topological charge for a fairly
large range in $m$. In figure 3 we show the correlation between this
quantity and the gluonic topological charge for the unimproved
staggered and for the fat7xasqtad actions. We can see that the two
definitions agree quite well for the improved action. In the
unimproved case there is still a correlation, but the two definitions
give very different results for the topological charge.

\begin{figure}
\includegraphics[scale=0.3, angle=-90]{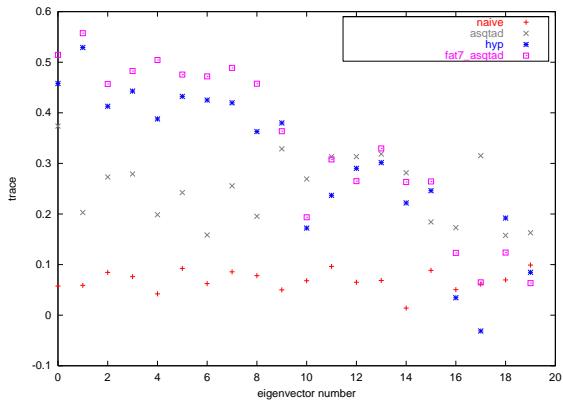}
\label{fig3}
\caption{Chirality of the lowest eigenvectors of the Dirac operator.}
\end{figure}

\begin{figure}
\includegraphics[scale=0.3, angle=-90]{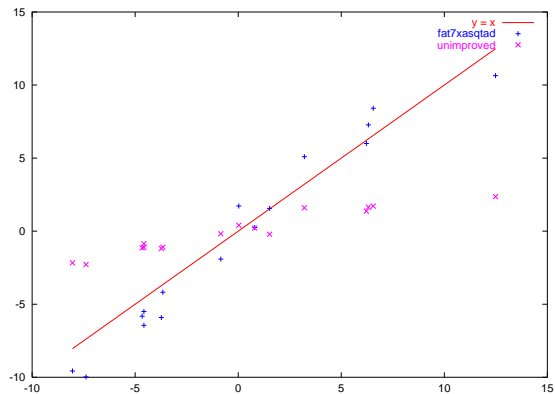}
\label{fig4}
\caption{Correlation between the gluonic and the fermionic definitions
of topological charge. The x symbols correspond to the unimproved
staggered action, whereas the + ones correspond to the fat7xasqtad
action. The straight line represents y = x.}
\end{figure}

\end{document}